\documentclass[prd,nofootinbib,preprintnumbers,preprint,
floatfix]{revtex4}
\usepackage{empheq}

\usepackage{amsmath}
\usepackage{amsfonts}
\usepackage{amssymb}
\usepackage{dsfont}
\usepackage{bm}
\usepackage[hyperfootnotes=false]{hyperref}
\usepackage[dvipsnames]{xcolor}
\usepackage{apptools}
\usepackage{enumerate}
\usepackage{longtable}

\newcounter{example}[section]
\newcounter{remark}[section]
\newcounter{theorem}[section]
\newcounter{proposition}[section]
\newcounter{lemma}[section]
\newcounter{corollary}[section]
\newcounter{definition}[section]

\AtAppendix{\counterwithin{example}{section}}
\AtAppendix{\counterwithin{remark}{section}}
\AtAppendix{\counterwithin{theorem}{section}}
\AtAppendix{\counterwithin{proposition}{section}}
\AtAppendix{\counterwithin{lemma}{section}}
\AtAppendix{\counterwithin{corollary}{section}}
\AtAppendix{\counterwithin{definition}{section}}

\def\theremark{\arabic{section}.\arabic{remark}}
\def\thetheorem{\arabic{section}.\arabic{theorem}}

\def\thedefinition{\arabic{section}.\arabic{definition}}

\renewcommand*{\email}[1]{\footnote{Electronic address: \href{mailto:#1}{\nolinkurl{#1}} }}

\begin{document}

\title{Perturbation theory for gravitational shadows in static spherically symmetric spacetimes}
\author{Kirill Kobialko${}^{1,\,}$\email{kobyalkokv@yandex.ru}}
\author{Dmitri Gal'tsov${}^{1,\,}$\email{galtsov@phys.msu.ru}}
\affiliation{${}^1$ Faculty of Physics, Moscow State University, 119899, Moscow, Russia}

\begin{abstract}
We develop a perturbation theory for surfaces confining photons and massive particles in static spherically symmetric spacetimes in terms of two parameters: the mass-to-energy ratio and the deviation of metric functions from a given form, e.g., the Schwarzschild solution. Expansions of the gravitational shadow radius in terms of these parameters are constructed up to the second order. The metric expansion in terms of the Schwarzschild mass-to-radius ratio is then reconstructed. Explicit analytical examples of non-standard black hole metrics are considered as an illustration. In some cases perturbative results demonstrate good accuracy even for non-small deviations.
\end{abstract}

\maketitle

\setcounter{page}{2}

\setcounter{equation}{0}
\setcounter{subsection}{0}

\section{Introduction}

Recent results of the Event Horizon Telescope collaboration on black hole shadows \cite{EventHorizonTelescope:2019dse,EventHorizonTelescope:2022wkp,Akiyama:2024zpm}
had a great impact on development of new theoretical approaches. It was realized that a number of phenomena in black hole physics are associated not only with the existence of an event horizon, but also with the special role of characteristic surfaces in strong gravitational fields on which photons and massive particles can remain infinitely. These surfaces are, in most astrophysically significant cases, unstable and serve as boundaries between streams of particles that scatter to infinity and are absorbed by black holes \cite{Virbhadra:1999nm,Virbhadra:2002ju,Virbhadra:2008ws,Shoom:2017ril,Gibbons:2016isj,Perlick:2021aok,Cunha:2018acu,Grenzebach:2014fha,Grenzebach:2015oea}.  Such surfaces are most simple for photons propagating in the static metrics, while in the rotating case similar role is played by surfaces where non-planar spherical orbits are located \cite{Teo:2020sey}. In the first case, the corresponding hypersurfaces in space-time are umbilic \cite{Claudel:2000yi} (the tensor of external curvature is proportional to the induced metric), in the second case they are partially umbilic \cite{Kobialko:2020vqf}, which means that the latter property is satisfied not for tensors as a whole, but by their convolutions with a part of the vectors of the tangent space. Equivalently, partially umbilic hypersurface can also be considered as totally umbilic in terms of the Jacobi metric \cite{Bermudez-Cardenas:2024bfi}. In a similar way, one can consider the characteristic surfaces of massive particles, as well as particles of variable mass, for example, photons in plasma \cite{Perlick:2015vta,Perlick:2017fio,Perlick:2023znh,Bezdekova:2022gib,Briozzo:2022mgg,Bogush:2023ojz,Kobialko:2022uzj,Song:2022fdg}. As in the case of photons in vacuum, this consideration leads to the consistent construction of analytical expressions for the so-called massive shadows  \cite{Kobialko:2023qzo} whose properties depend on the particles energy. 

Besides the direct application to strong gravitational lensing, characteristic surfaces are of key importance in the analysis of a number of mathematical problems of black hole physics. It was noted \cite{Pappas:2018opz,Glampedakis:2018blj,Konoplya:2021slg} that the existence of hypersurfaces with the above properties in the general stationary case correlates with the existence of hidden symmetries for spacetimes exhibited by Killing tensors. 
Eventually, it was shown \cite{Kobialko:2021aqg,Kobialko:2022ozq} that Killing tensors, which reduce to trivial (direct products of Killing vectors) on hypersurfaces foliating all of spacetime, ensure that these hypersurfaces contain generalized photon surfaces, including those associated with the motion of massive particles.
This in turn entails integrability of Einstein equations \cite{Galtsov:2024vqo}. The existence of photon and massive particles spheres allows to formulate a set of novel uniqueness theorems for black holes \cite{Cederbaum:2019rbv,Yazadjiev:2015hda,Yazadjiev:2015jza,Rogatko:2016mho,Kobialko:2024rqr}.

Recently the list of experimentally observed images increased greatly which stimulated search of new theoretical methods of their description and classification \cite{Vagnozzi:2022moj}. Forty strong lenses were analysed in \cite{Virbhadra:2022ybp,Virbhadra:2022iiy} and a new concepts of compactness and distortions was introduced. However, in the general case, within the framework of the classification and analytical analysis of strong gravitational lensing, the problem of the inaccessibility of fully analytical expressions arises. In this case, it is necessary to resort to a numerical method, which complicates the task of analyzing families of gravitational models. Here we develop perturbational technique which seems useful for further analysis of shadows images. This approach clarifies the role of massive particle surfaces (MPS) and their relation with more familiar photon surfaces and also opens the way to analize possible deviations of black hole metrics from the standard Schwarzschild metric using shadows.
Here we restrict by the simplest case of static spherically symmetric metrics. A similar problem was considered recently in \cite{Vertogradov:2024dpa}.

The article is organized as follows: Section \ref{sec:massive_shadows} contains the derivation of the general analytical formulas for the radii of MPS and gravitational shadow. Section \ref{sec:shadows_expansion} presents a step-by-step construction of two parameters perturbation theory for the MPS (first order inclusive) and the massive shadow (second order inclusive). In Section \ref{sec:metric_reconstruction}, a new algorithm for restoring metric parameters from a set of massive shadows is proposed. Section \ref{sec:examples} contains a set of examples aimed at confirming the correctness of the obtained expressions, as well as at obtaining new results and comparing predictions in different gravity models. Conclusion \ref{sec:conclusion} contains a brief summary of the main results and their discussion.

\section{Massive shadows} \label{sec:massive_shadows}

Our first goal is to obtain an explicit analytical expression describing the boundary of the gravitational shadow arising in streams of massive and massless particles scattering on static spherically symmetric black holes or other ultracompact objects. Since here we are not interested in general theoretical questions such as uniqueness theorems \cite{Kobialko:2024rqr} etc., we prefer to follow the direct geodesic approach instead of the massive particle surface formalism \cite{Kobialko:2022uzj}. Equivalence of both approaches was recently demonstrated in \cite{Bogush:2024fqj}.

Consider a general static spherically symmetric four-dimensional spacetime with the following metric tensor \cite{Perlick:2021aok,Vagnozzi:2022moj}
\begin{equation}\label{eq:metric_tensor}
    ds^2 = -\alpha dt^2 + \gamma dr^2 + \beta \left(d\theta^2 + \sin^2\theta d\phi^2\right),
\end{equation}
where $\alpha, \beta, \gamma$ are functions of $r$ and some parameter $\delta$, i.e. $\alpha=\alpha(r,\delta)$ and so on. As parameter $\delta$ one can choose ADM mass $M$, electric $Q$ or scalar charge, etc. The metric (\ref{eq:metric_tensor}) has two obvious Killing vectors $\partial_t$ and $\partial_\phi$ which correspond to two conservation laws. Moreover, spherical symmetry allows us to limit our consideration to equatorial motion $\theta=\pi/2$. 

Consider some timelike or null equatorial geodesic $x^\alpha(s)$. The conservation laws and normalization conditions for such a geodesic take the form
\begin{equation} \label{eq:laws}
\dot{x}^\alpha \dot{x}_\alpha=-m^2, \quad \alpha \dot{x}^t = E, \quad  \beta\dot{x}^\phi = L, \quad \dot{x}^\alpha=\frac{dx^\alpha(s)}{ds},
\end{equation}
where $m$, $E$ and $L$ are particle mass, conserved energy and angular momentum respectively. After explicit substitution into the normalization condition the conservation laws and the condition of equatorial motion $\dot{x}^\theta=0$ we find
\begin{equation} 
  E^{-2}\gamma \cdot(\dot{x}^r)^2 =V\equiv \alpha^{-1} -\beta^{-1}  l-\epsilon, \quad \epsilon=m^2/E^2, \quad l=  L^2/E^2,
\end{equation}
where we introduced the effective potential $V=V(r,\delta)$. 

It is well known that the boundary of the gravitational shadow is formed by geodesics passing through the observation point and asymptotically winding onto the photon sphere or massive particles sphere \cite{Perlick:2021aok,Cunha:2018acu,Grenzebach:2014fha,Grenzebach:2015oea,Kobialko:2023qzo}. These spheres represent a union of circular orbits $r=const$ and can be defined by the common solution of the equations $V=0$ and $V_{,r}=0$, where comma with the index $r$ means partial derivative with respect to $r$. For our choice of effective potential we get  
\begin{equation} \label{eq:equations}
  \epsilon=\alpha^{-1} -\beta^{-1}  l,\quad
  -\frac{\alpha_{,r}}{\alpha^{2}} +\frac{\beta_{,r}}{\beta^{2}} \cdot l=0,
\end{equation}
from which we immediately obtain
\begin{equation}
  \epsilon =\alpha^{-1}\left(1-\frac{\beta}{\alpha}\cdot\frac{\alpha_{,r}}{\beta_{,r}}\right), \quad l=\frac{\beta^{2}}{\alpha^{2}}\cdot\frac{\alpha_{,r}}{\beta_{,r}}.
\end{equation}
To determine the stability of geodesics on the MPS, it is also necessary to calculate the second derivative of the effective potential (stable if $V_{,rr}<0$, unstable if $V_{,rr}>0$) 
\begin{equation} \label{eq:V_rr}
V_{,rr}=\frac{\beta_{,rr}}{\alpha^{2}}\cdot\frac{\alpha_{,r}}{\beta_{,r}} -\frac{\alpha_{,rr}}{\alpha^{2}}+\frac{2\alpha_{,r}}{\alpha^{2}}\cdot\left(\frac{\alpha_{,r}}{\alpha}  -\frac{\beta_{,r}}{\beta}  \right)=\frac{\beta_{,r}}{\beta}\cdot \epsilon_{,r}.
\end{equation}
Thus, the (un)stability of the MPS can be determined by the sign of the derivative of energy $\epsilon_{,r}$ along the flow of slices $r=const$, which was established earlier in Ref. \cite{Kobialko:2022uzj}. In fact, by changing the coordinates, we can always ensure that $\frac{\beta_{,r}}{\beta}>0$ is fulfilled what makes the connection unambiguous. Moreover, from the properties of the inverse function we have 
\begin{equation} \label{eq:r_epsilon}
r_{,\epsilon}=\left(\epsilon_{,r}\right)^{-1}=\frac{\beta_{,r}}{\beta V_{,rr}}.
\end{equation}
Therefore, the radius (in coordinates such that $\frac{\beta_{,r}}{\beta}>0$) of the unstable MPS of a static spherically symmetric spacetime increases, i.e $r_{,\epsilon}>0$. Since the shadow is usually formed as a result of winding geodesics on the MPS, it is natural to expect that such surfaces should be unstable. So we can naturally assume that the radius of the MPS on which scattering occurs actually increases with $\epsilon$.

At the observation point $\bar{r}$ for a static equatorial ($\theta=\pi/2$) observer we have the following expression for the tangent vector of the observed geodesic (a standard orthonormal tetrad is chosen \cite{Perlick:2021aok,Kobialko:2023qzo})
\begin{equation} \label{eq:orthonormal_tetrad}
\dot{x}^t=\frac{a}{\sqrt{\bar{\alpha}}} , \quad \dot{x}^r = \frac{b}{\sqrt{\bar{\gamma}}} \cos \Theta, \quad  \dot{x}^\phi =  \frac{b}{\sqrt{\bar{\beta}}} \sin \Theta, 
\end{equation}
where $\Theta$ is an angle on the observer's celestial sphere between the north pole and the direction of the geodesic detection, $a$ and $b$ are some constants. The bar means that all functions are calculated at the observation point, i.e. $\bar{\alpha}=\alpha(\bar{r},\delta)$. Since space is spherically symmetric, the shadow will be a circle on celestial sphere with constant $\Theta=const$. After explicit substitution of (\ref{eq:orthonormal_tetrad}) in (\ref{eq:laws}) calculated at the observation point we find 
\begin{equation}
-m^2 = - a^2 + b^2, \quad \sqrt{\bar{\alpha}} a=E, \quad \sqrt{\bar{\beta}} b \sin \Theta =L. 
\end{equation}
Solving this system with respect to $\sin \Theta$ we get
\begin{equation}
\sin^2 \Theta =\frac{\bar{\alpha}}{\bar{\beta}}\cdot\frac{l}{1-\bar{\alpha}\epsilon}.
\end{equation}
Substituting $l$ from (\ref{eq:equations}) for angular size of the gravitational shadow we get the general expression
\begin{align} \label{eq:a1}
\sin^2\Theta = \frac{\bar{\alpha}}{\alpha}\cdot \frac{\beta}{\bar{\beta}}\cdot \frac{1-\alpha \epsilon}{1-\bar{\alpha}\epsilon}. 
\end{align}
Note that an alternative description of the shadow can be given using a stereographic projection of the celestial sphere. The radius of the shadow in the stereographic projection coordinates ($X,Y$) reads as \cite{Kobialko:2023qzo,Bogush:2022hop} 
\begin{align}
R_{proj}=\sqrt{X^2+Y^2}= 2\tanh \left(\frac{\Theta}{2}\right).
\end{align}
In what follows, we will be primarily interested in the shadows observed by an asymptotic observer
$\bar{r}\rightarrow \infty$ in asymptotically flat spacetimes, i.e.
\begin{align}
\bar{\alpha}=1, \quad \bar{\beta}=\bar{r}^2.
\end{align}
In this case, the shadow size tends to zero and
\begin{align} \label{eq:a2}
R^2_{proj}=\Theta^2=\frac{1}{\bar{r}^2}\cdot \frac{\beta}{\alpha}\cdot \frac{1-\alpha \epsilon}{1-\epsilon}.
\end{align}
For convenience, we introduce the finite shadow radius as \cite{Perlick:2021aok} 
\begin{align} \label{eq:def1}
R\equiv \lim_{\bar{r}\to \infty}(\bar{r}\cdot R_{proj}).
\end{align}
Then from (\ref{eq:a2}) we find a particularly simple expression for the square of shadow radius
\begin{align} \label{eq:area_shadow}
R^2=  \frac{\beta}{\alpha}\cdot \frac{1-\alpha \epsilon}{1-\epsilon}, \quad 
   \epsilon =\frac{1}{\alpha}\left(1-\frac{\beta}{\alpha}\cdot\frac{\alpha_{,r}}{\beta_{,r}}\right).
\end{align}
In the case of null geodesics we find
\begin{align}
R^2=  \frac{\beta}{\alpha}, \quad 
  \beta\alpha_{,r}=\alpha\beta_{,r}.
\end{align}
This result is consistent with the result of Ref. \cite{Perlick:2021aok} (eq. (22) and (27)). 

\section{Shadows expansion} \label{sec:shadows_expansion} 

The equations (\ref{eq:area_shadow}) obtained in the previous section define a parametric representation of the family of shadows created by particles with different energies. Unfortunately, the parameter of the family is the radius of the MPS, a value that is practically not measurable experimentally. In fact, it would be much more convenient for us to use $\epsilon$ itself as a parameter, which will make it easy to isolate the massless case and fit experimental data. However, the problem is that even in simple cases functional dependency of MPS radius $r=r(\epsilon,\delta)$ can only be obtained as a solution of a high-degree polynomial equation. Of course, this problem can be solved numerically, for example using Newton's method. However, it is also useful to develop a perturbation theory for this problem in two directions: one is to consider light particles with high energies such as neutrinos and another to consider metrics deviating from Schwarzschild. A similar approach was recently  applied to the massless case \cite{Vertogradov:2024dpa}.

We will construct series expansion of the general shadow radius  (\ref{eq:area_shadow}) in terms of the parameters $\epsilon$ and $\delta$ with an accuracy up to the second order. In these calculations, the second equation in (\ref{eq:area_shadow}) should be viewed as an implicit assignment of the MPS radius $r=r(\epsilon,\delta)$, which is then substituted into the first equation in (\ref{eq:area_shadow}).
As we will see later, for the shadow decomposition up to the second order it is sufficient to use the expansion of $r$ only up to the first order. 

\subsection{Derivatives} 

We start with a slightly more general problem, namely, calculating the derivatives of $R^2$ (\ref{eq:area_shadow}) with respect to $\epsilon$ and $\delta$ at an arbitrary point. Direct calculations give 
\begin{align} \label{eq:inter1}
R^2_{,\epsilon}= \frac{\beta}{\alpha}\cdot\left[ \frac{1-\alpha }{ (1-\epsilon)^2 }-\frac{\beta  \alpha_{,r} -\alpha  \beta_{,r}  (1-\alpha  \epsilon )}{\alpha  \beta  (1- \epsilon )}\cdot r_{,\epsilon}\right], 
\end{align}
where $r_{,\epsilon}$ was defined in equation (\ref{eq:r_epsilon}). 
On the other hand, using the second equation in (\ref{eq:area_shadow}), for the numerator of the second term in (\ref{eq:inter1}), we get
\begin{align}
\beta  \alpha_{,r} -\alpha  \beta_{,r}  (1-\alpha  \epsilon )=\beta  \alpha_{,r} -\alpha  \beta_{,r} \left[1- \left(1-\frac{\beta}{\alpha}\cdot\frac{\alpha_{,r}}{\beta_{,r}}\right)\right]=0.
\end{align}
Therefore the multiplier before $r_{,\epsilon}$ becomes zero on the MPS. Thus, the first order derivative of $R^2$  formally does not depend on the MPS radius $r$ derivative at all and we find 
\begin{align} \label{eq:R_epsilon}
R^2_{,\epsilon}= \frac{\beta}{\alpha}\cdot\frac{1-\alpha }{ (1-\epsilon)^2 }.
\end{align}
Proceeding in exactly the same way to calculate the derivative of $R^2$  with respect to $\delta$, we obtain
\begin{align}\label{eq:R_delta}
R^2_{,\delta}=-\frac{\beta }{\alpha}\cdot\frac{ \frac{\alpha_{,\delta} }{\alpha }-(1-\alpha  \epsilon )\frac{\beta_{,\delta}  }{\beta }}{  1-\epsilon }.
\end{align}
Note that, as in the case of $\epsilon$ derivative, the multiplier before $r_{,\delta}$ becomes zero on the MPS (the multiplier is absolutely identical). Moreover, such a multiplier will be retained in the higher order derivatives. In particular, to calculate the second order derivatives of the shadow square radius, we need only the first order derivatives of the MPS radius $r$. It should be noted that such a pattern only occurs when the shadow radius is given by the expression (\ref{eq:area_shadow}). If we make an explicit substitution of $\epsilon$ or use the photon surface equation as was done in Ref. \cite{Vertogradov:2024dpa}, the expressions will become formally dependent on the first order derivatives of the MPS radius. But, after all substitutions, the results must be identical.

One of first-order derivatives of the MPS radius $r$ was found earlier (\ref{eq:r_epsilon}), while the other one has a more complex form 
\begin{align}  \label{eq:r_delta}
r_{,\delta}=\frac{-\beta   \alpha_{,\delta }  \beta_{,r}  (1-2 \alpha  \epsilon )+\beta^2 \alpha_{,r\delta}+\alpha  \left(\beta_{,\delta} \beta_{,r}-\beta  \beta_{,r\delta}\right) (1-\alpha  \epsilon)}{\alpha^2 \beta^2 V_{,rr}}.
\end{align}
With these preparations, after some extensive calculations for second order derivatives we obtain 
\begin{align}
R^2_{,\epsilon\epsilon}&=\frac{2\beta }{\alpha   }\frac{1- \alpha}{ (1-\epsilon )^3} -\frac{\beta V_{,rr} }{(1-\epsilon )}\cdot r^2_{,\epsilon},\label{eq:R_epsilon_epsilon}\\
R^2_{,\delta\delta}&=\frac{  \alpha  \beta_{,\delta\delta} (1-\alpha  \epsilon )-2 \alpha_{,\delta} \beta_{,\delta}-\beta  \left(  \alpha_{,\delta\delta}-2 \alpha^{-1}\alpha_{,\delta}^2\right)}{\alpha^2 (1-\epsilon )}-\frac{ \beta V_{,rr} }{ (1-\epsilon )}\cdot r^2_{,\delta},\label{eq:R_delta_delta}\\R^2_{,\epsilon\delta}&=
\frac{ (1-\alpha) \alpha  \beta_{,\delta}-\beta  \alpha_{,\delta}}{\alpha^2  (1-\epsilon)^2}-\frac{\beta V_{,rr} }{(1-\epsilon )}\cdot r_{,\epsilon}r_{,\delta}.\label{eq:R_epsilon_delta}
\end{align}
These derivatives allow us to find the following general shadow radius expansion up to the second order around any point ($\epsilon_0,\delta_0$):
\begin{align} \label{eq:expansion}
R^2(\epsilon,\delta)=&R^2(\epsilon_0,\delta_0)+R^2_{,\epsilon}(\epsilon_0,\delta_0)\cdot(\epsilon-\epsilon_0)+R^2_{,\delta}(\epsilon_0,\delta_0)\cdot(\delta-\delta_0)\\&+R^2_{,\epsilon\epsilon}(\epsilon_0,\delta_0)\cdot\frac{(\epsilon-\epsilon_0)^2}{2}+R^2_{,\delta\delta}(\epsilon_0,\delta_0)\cdot\frac{(\delta-\delta_0)^2}{2}+R^2_{,\epsilon\delta}(\epsilon_0,\delta_0)\cdot(\epsilon-\epsilon_0)(\delta-\delta_0)+...\nonumber
\end{align}
As a point over which the expansion is performed, we will choose the well-known photon sphere/shadow or MPS for which explicit analytical expressions are known. 

\subsection{$\epsilon$-expansion over photon sphere}
In the zero order of $\epsilon$ expansion we obviously arrive at the usual photon sphere ($\epsilon=0$) whose radius $r_{PS}$ is determined from the equation (which can often be resolved in radicals) \cite{Perlick:2021aok,Vagnozzi:2022moj} 
\begin{align} \label{eq:photon_sphere}
\left[\frac{\alpha_{,r}}{\alpha}-\frac{\beta_{,r}}{\beta}\right]_{r_{PS}}=0.
\end{align}
To calculate the first order perturbation of photon sphere, we will use the formula (\ref{eq:r_epsilon}) for the first derivative of $r$ obtained earlier.
\begin{align}
 r_{,\epsilon}\Big|_{r_{PS}}=\frac{\beta_{,r}}{\beta V_{,rr}}\Big|_{r_{PS}},
\end{align}
where by virtue of the photon sphere equation (\ref{eq:photon_sphere}) expression (\ref{eq:V_rr}) takes the form 
\begin{align}
V_{,rr}\Big|_{r_{PS}}=\left[\frac{\beta_{,rr}}{\alpha \beta} -\frac{\alpha_{,rr}}{\alpha^{2}}\right]_{r_{PS}}.
\end{align}
The full first order expansion for $r$ is
\begin{align} \label{eq:epsilon_expansion_r}
r = r_{PS} +\frac{\beta_{,r}}{\beta V_{,rr}}\Big|_{r_{PS}} \cdot \epsilon +O(\epsilon).
\end{align}
The expression for the square radius of the shadow in zero perturbation order is ($\epsilon=0$)
\begin{align} \label{eq:a9}
R^2_{PS} = \frac{\beta}{\alpha}\Big|_{r_{PS}}.
\end{align}
The first order perturbation is determined by the derivative (\ref{eq:R_epsilon}):
\begin{align}
R^2_{,\epsilon}\Big|_{r_{PS}}=R^2_{PS}\cdot(1-\alpha)\Big|_{r_{PS}}.
\end{align}
The second order perturbation is determined by the derivative (\ref{eq:R_epsilon_epsilon}):
\begin{align}
R^2_{,\epsilon\epsilon}\Big|_{r_{PS}}=2R^2_{PS}\cdot(1-\alpha)\Big|_{r_{PS}} -\beta V_{,rr}\cdot r^2_{,\epsilon}\Big|_{r_{PS}}.
\end{align}
Collecting these expressions together and using the Eq. (\ref{eq:expansion}) we get the following $\epsilon$-expansion over the photon sphere
\begin{align} \label{eq:epsilon_expansion}
R^2=R^2_{PS}\cdot \left[1+(1-\alpha)\cdot\epsilon+\left(1-\alpha-\frac{1}{2}\cdot\frac{\alpha^2}{\beta^2}\cdot\frac{\beta^2_{,r} }{\beta^{-1}\beta_{,rr} -\alpha^{-1}\alpha_{,rr}} \right)\cdot \epsilon^2\right]_{r_{PS}} +O(\epsilon^2). 
\end{align}
This expansion should be used in cases where the expression for a photon sphere is well defined analytically, while determining the radius of the MPS for a given $\epsilon\neq0$ is problematic.

\subsection{$\delta$-expansion over Schwarzschild metric}

Our current goal is to obtain a series expansion of general shadow square radius (\ref{eq:area_shadow}) in parameter $\delta$ up to the second order over Schwarzschild metric. As we will see, unlike most metrics, Schwarzschild solution allows for the simplest expression for the radius of the MPS and, as a consequence, the massive shadow radius. Thus, it is a natural point of decomposition. We will also choose as $r$ the area radius of $r=const$ sphere, i.e. $\beta=r^2$. This choice of coordinates does not violate generality, although it may lead to more complicated calculations for some metrics where the natural coordinates are different. Thus, the components of the metric are represented up to the second order in the form 
\begin{align} \label{eq:a3}
\alpha=1-\frac{2M}{r} +\alpha_{1}(r/M) \cdot \delta + \alpha_{2}(r/M) \cdot\delta^2, \quad \beta=r^2.
\end{align}
In the zero order of perturbation, the radius of the MPS is determined from the second equation in (\ref{eq:area_shadow}) with $\delta=0$
\begin{align} \label{eq:a4}
\epsilon=\frac{r^2_{Sch}-3 M r_{Sch}}{(r_{Sch}-2M)^2},
\end{align}
and reads as \cite{Kobialko:2022uzj} (we drop the extra negative root)
\begin{align} \label{eq:a5}
r_{Sch}=M f, \quad f=\frac{3 -4 \epsilon+\sqrt{9-8\epsilon}}{2(1-\epsilon)}, \quad 0\leq\epsilon<1.
\end{align}
For the perturbation of the MPS in the first order (\ref{eq:r_delta}) we find 
\begin{align}
r/M=f+\frac{f^2 \left[(f-2) f\cdot \alpha'_1(f)-2 (4-f)\cdot \alpha_1(f)\right]}{2 (6-f)}\cdot \delta+...,
\end{align}
where we make an explicit substitution of $\epsilon$ at the end because this simplifies the final expressions. 

The expression for the massive shadow square radius in zero order can be obtained from (\ref{eq:area_shadow}) by substituting (\ref{eq:a3}) with $\delta=0$:
\begin{align} \label{eq:SH_shadow}
R^2_{MSch}=M^2\cdot\frac{f^3 }{4-f},
\end{align}
where we made an explicit substitution of $\epsilon$ from (\ref{eq:a4}) again. This is nothing than an expression for massive shadow square radius in Schwarzschild metric. 

In the first order of perturbation expansion (\ref{eq:R_delta}) we get
\begin{align}
R^2_{,\delta}\Big|_{r_{Sch}}=-M^2\cdot\frac{f^4 }{4-f}\cdot\alpha_1(f).
\end{align}
In the second order (\ref{eq:R_delta_delta})
\begin{align}
R^2_{,\delta\delta}\Big|_{r_{Sch}}=&-M^2\cdot\frac{f^4}{2(6-f)(4-f)}\cdot\Big[(f-2) f^3 \alpha'_1(f)^2+4 (f-4) f^2 \alpha_1(f) \alpha'_1(f)\nonumber\\&+4 (f-5) f \alpha_1(f)^2-4 (f-6) \alpha_2(f)\Big].
\end{align}
Thus we get the following $\delta$-expansion over Schwarzschild metric
\begin{align} \label{eq:delta_expansion}
R^2=&R^2_{MSch}\cdot \Big[1-f\alpha_1(f)\cdot \delta-\frac{f}{4(6-f)}\cdot\Big\{(f-2) f^3 \cdot\alpha'_1(f)^2-4 (4-f) f^2 \cdot\alpha_1(f) \alpha'_1(f)\nonumber\\&-4 (5-f) f\cdot \alpha_1(f)^2+4 (6-f) \cdot\alpha_2(f)\Big\}\cdot \delta^2\Big] +O(\delta^2),
\end{align}
where $f$ is defined explicitly by (\ref{eq:a5}). This expression is useful in all cases where the metric of interest is a perturbation of the Schwarzschild metric. In particular, it is applicable to the analysis of perturbations of photon shadows.

\subsection{$\delta,\epsilon$-expansion over Schwarzschild photon sphere} 

Finally, consider a case that is at the intersection of the previous two, namely, low mass/high energy particles and weak deviations from the Schwarzschild simultaneously. The metric is represented up to the second order also in the form (\ref{eq:a3}).

At zero order we simply get a photon sphere in the Schwarzschild metric $r_{PS}=3M$. In the first order, the perturbation of the photon sphere has the form (\ref{eq:r_epsilon}) and (\ref{eq:r_delta}) with substitution $r=3M$, $\epsilon=0$, $\delta=0$:
\begin{align} \label{eq:b4}
r = 3M + \frac{M}{3}\cdot \epsilon +M \cdot \left[\frac{9}{2}\cdot \alpha_1'(3)-3 \cdot\alpha_1(3 )\right]\cdot \delta +O(\epsilon,\delta).
\end{align}
The expression for the shadow square radius in zero order is 
\begin{align}
R^2_{Sch} = 27M^2.
\end{align}
This result is consistent with \cite{Perlick:2021aok}.

In the first order from (\ref{eq:R_epsilon}) and (\ref{eq:R_delta}) we get
\begin{align}
R^2_{,\epsilon}\Big|_{r_{PS}}= 18M^2,\quad R^2_{,\delta}\Big|_{r_{PS}}=- 81M^2 \cdot\alpha_1(3).
\end{align}
In the second order from (\ref{eq:R_epsilon_epsilon}), (\ref{eq:R_delta_delta}) and (\ref{eq:R_epsilon_delta}) we get
\begin{align}
R^2_{,\epsilon\epsilon}\Big|_{r_{PS}}&= 34M^2, \\
R^2_{,\epsilon\delta}\Big|_{r_{PS}}&= - 9\left[3 \alpha_1'(3)+7 \alpha_1(3)\right]\cdot M^2,\\
R^2_{,\delta\delta}\Big|_{r_{PS}}&=  -\frac{81}{2}\left[9\alpha_1'(3)^2-12 \alpha_1(3) \alpha'_1(3)-8 \alpha_1(3)^2+4 \alpha_2(3)\right]\cdot M^2.
\end{align}
Thus we end up with the following $\delta,\epsilon$-expansion over Schwarzschild photon sphere
\begin{align} \label{eq:a6}
R^2=& R^2_{Sch} \cdot\Big[1+\frac{2}{3}\cdot \epsilon-3\alpha_1(3 )\cdot\delta+\frac{17}{27}\cdot\epsilon^2-\left[\alpha_1'(3 )+\frac{7}{3}\alpha_1(3)\right]\cdot\epsilon \delta+\nonumber\\&+\left[-\frac{27}{4} \alpha_1'(3)^2+9  \alpha_1(3 ) \alpha_1'(3 )+6 \alpha_1(3 )^2-3 \alpha_2(3 )\right]\cdot \delta^2\Big]+O(\epsilon^2,\delta^2,\epsilon \delta).
\end{align}
This formula has a simpler form than (\ref{eq:delta_expansion}) and allows one to analyze small deviations from Schwarzschild using high-energy particles.

\section{Metric reconstruction} \label{sec:metric_reconstruction}

An important special case of the formulas (\ref{eq:delta_expansion}) or (\ref{eq:a6}) is the case of perturbation of the photon shadow ($\epsilon=0$) in the first order, which intersects with Ref. \cite{Vertogradov:2024dpa}. Our result in this case is given by the following expression  
\begin{align}
R^2= 27 M^2 -81M^2\alpha_1(3)\cdot\delta+O(\delta).
\end{align}
If we use an explicit form of $\alpha$ similar to that proposed in the Ref. \cite{Vertogradov:2024dpa} 
\begin{align} \label{eq:a31}
\alpha=1-\frac{2M}{r} +\sum^n_{i=2} \frac{c_i}{r^i},
\end{align}
i.e
\begin{align}
\alpha_1(r/M) \cdot\delta= \sum^n_{i=2} \frac{c_i}{r^i}=\sum^n_{i=2}(c_i/M^i)\cdot \frac{M^i}{r^i}.
\end{align}
we obtain very simple expansions for photon spheres 
\begin{align} \label{eq:f1}
r_{PS} = 3M -\frac{1}{2} \cdot \sum^n_{i=2} \frac{(i+2)c_i}{(3M)^{i-1}}+O(c_i),
\end{align}
and for photon shadow $R^2_{PS}$ 
\begin{align} \label{eq:a7}
R^2_{PS}= 27 M^2 -81 M^2\cdot\sum^n_{i=2} \frac{c_i}{(3M)^i}+O(c_i).
\end{align}
An important feature of the obtained formula (\ref{eq:a7}) is that it gives an expression for the square of the photon shadow radius as a linear combination of coefficients $c_i$. Thus, if we have only one such parameter, we can restore this parameter unambiguously by the measurement of the shadow size. At the same time, it is impossible to uniquely restore the coefficients separately if there are many of them since we lack equations. However, massive particles provide such an equations.

Keeping the same notations (\ref{eq:a31}) and retaining the first order of corrections in (\ref{eq:delta_expansion}) for the square of the massive shadow radius we find  
\begin{align} 
R^2= M^2\cdot\frac{f^3 }{4-f} \left(1-f \sum^n_{i=2}\frac{c_i}{(Mf)^i}\right)+O(c_i),
\end{align}
where $f$ is defined in (\ref{eq:a5}). Suppose that we have made several measurements of the shadow radius $R_j$ for particles with a set of $\epsilon_j$. Then we obtain a system of non-homogeneous linear equations (subsequent equations are obtained by discarding the corrections $O(c_i)$)
\begin{align}\label{c1}
R^2_j= M^2\cdot\frac{f^3_j }{4-f_j} \left(1-f_j \sum^n_{i=2}\frac{c_i}{(Mf_j)^i}\right),  \quad j=2,...,n,
\end{align}
and we reserved index $j=1$ for the photon shadow (\ref{eq:a7}). By selecting enough measurements of $\epsilon_j$ and $R_j$ , we can, in principle, express the coefficients $c_i$ uniquely as a solution of this system. However, these solutions contain problematic to measure distance to the black hole $\bar{r}$ (see definition (\ref{eq:def1})) and its mass $M$. Let us therefore consider the ratios  
\begin{align} 
\chi_j=(R_j/R_{1})^2, \quad R_{1}=R_{PS}.
\end{align}
From equation (\ref{eq:def1}) it is clear that parameter $\bar{r}$ excluded in these ratios. Then combining the equations (\ref{eq:a7}) and (\ref{c1}) we get 
\begin{align} \label{c2}
27(4-f_j)\chi_j-f^3_j= \sum^n_{i=2} \left(3^{4-i} \left(4-f_j\right)\chi_j-f^{4-i}_j \right)\cdot \frac{c_i}{M^i}.
\end{align}
Let's define 
\begin{align} 
N_j^i=3^{4-i}\left(4-f_j\right) \chi_j-f^{4-i}_j, \quad C_i=c_i/M^i.
\end{align}
Then we can write the equations (\ref{c2}) in the form $n-1$ non-homogeneous
linear equations for $n-1$ unknown coefficients
\begin{align}  \label{c3}
 \sum^n_{i=2}N_j^i C_i=N_j^1, \quad j=2,...,n.
\end{align}
Thus, if we can make enough measurements of the massive shadow radius for different energies, we can reconstruct the coefficients of the metric by solving an inhomogeneous system of linear equations (\ref{c3}). Moreover, we do not need any data about the distance to the black hole or its mass. Although of course knowledge of the mass is necessary to determine the absolute value for $c_i=C_i M^i$. This result looks encouraging since it requires determining only the quantities observed on Earth directly.    

Note also an important problems of the approaches. We can choose the number of $n$ manually as well as we can set a priori for some $c_i=0$ and consider fewer equations. The result may generally depend on this. Thus, the problem of determining the coefficients can be solved accurately only if we establish which coefficients $c_i$ are different from zero, i.e. a specific gravity model. If the theory is not known, the following criterion can be proposed: If $c_i\neq0$ is chosen correctly, then choosing another sequence of energies should yield the same result. In this paper we will consider only the application of this method to determine parameters in a priori given models and leave the issue of determining models from a set of shadows for future research. In addition, from the expansion (\ref{eq:a6}) it becomes clear that in the first order in $\epsilon$ expansion it is possible to obtain only one unique equation, while the remaining equations arise from higher orders. Therefore, the greatest accuracy of determining $c_i$ can be achieved by choosing a larger spread for $\epsilon_j$ which could be problematic for a potential experiment. However, of course this problem does not arise in the case where only one $c_i$ is different from zero. 

There is also an experimental problem of detecting a massive shadow. One could propose neutrinos as such scattered particles, since their motion from the black hole to the Earth will presumably be weakly affected. However, these particles are very difficult to detect and measure their energy-to-mass ratio. Therefore, observing a shadow in neutrino beams looks like a very distant prospect. Another way to see a massive shadow may be related to a secondary electromagnetic radiation associated with the scattering of massive particles.

\section{Examples}  \label{sec:examples}

\subsection{Schwarzschild metric}

\begin{figure}[tb]
\centering
\includegraphics[scale=0.7]{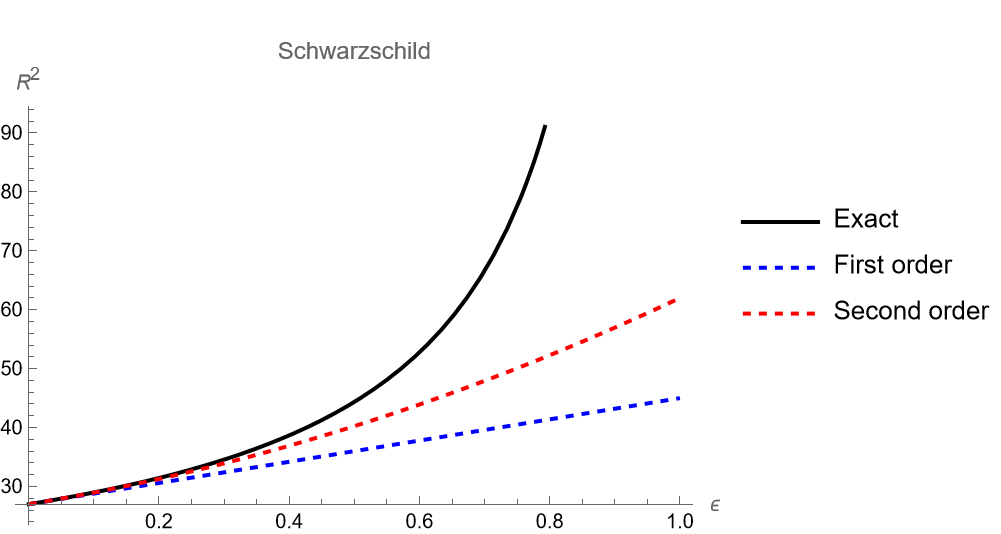} 
		\caption{Schwarzschild massive shadow. The dependence of the massive shadow square radius $R^2$ on $\epsilon$ ($M=1$). The exact result, first-order approximation, second-order approximation are presented.}
		\label{fig:SH}
\end{figure}

We are interested in the following Schwarzschild metric components \cite{Stephani:2003tm}
\begin{align}
\alpha=1-\frac{2M}{r}, \quad \beta=r^2.
\end{align}
As described earlier (\ref{eq:SH_shadow}), the exact expression for the square of the shadow radius is
\begin{align}
R^2=M^2\cdot\frac{f^3 }{4-f  }, \quad f=\frac{3 -4 \epsilon+\sqrt{9-8\epsilon}}{2(1-\epsilon)}.
\end{align}
If we expand this expression into a series in $\epsilon$, we get
\begin{align}
R^2= 27M^2\cdot \left(1+\frac{2}{3}\cdot\epsilon+\frac{17}{27}\cdot\epsilon^2\right)+...
\end{align}
It is easy to see that the first terms of the expansion exactly reproduce the result (\ref{eq:a6}) for $\delta=0$ as well as the result (\ref{eq:epsilon_expansion}) for $r_{PS}=3M$. In special case $\epsilon=0$ we find $R^2_{Sch}= 27M^2$ \cite{Perlick:2021aok,Vagnozzi:2022moj}. A graphical illustration of the approximations is shown in the Fig. \ref{fig:SH}.

\subsection{Reissner-Nordström metric}

The Reissner-Nordström metric is \cite{Stephani:2003tm}
\begin{align}
\alpha=1-\frac{2M}{r} +\frac{M^2\delta}{r^2}, \quad \beta=r^2, \quad \delta=Q^2/M^2.
\end{align}
The radius of the MPS is determined from the second equation in (\ref{eq:area_shadow}) which reads as  \cite{Kobialko:2022uzj}
\begin{align} \label{eq:b1}
\epsilon=\frac{r^2 \left(r (r-3 M)+2 M^2\delta\right)}{\left(r (r-2 M)+M^2\delta\right)^2}, 
\end{align}
and corresponds to roots of a fourth-degree polynomial. Although such a root can be expressed in radicals, it looks extremely cumbersome. However, we can find the series expansion of the root in the neighborhood of $r=3M$:
\begin{align} \label{eq:b2}
r=3 M +\frac{M}{3}\cdot \epsilon - \frac{2M}{3}\cdot \delta+ \frac{5M}{27}\cdot \epsilon^2-\frac{2M}{27}\cdot \epsilon \delta-\frac{4M}{27}\cdot \delta^2+...
\end{align}
The equation (\ref{eq:b1}) is simplified for the case of a photon sphere $\epsilon=0$. In particular, the radius is easily expressed through radicals (we choose one of the 2 roots which corresponds to the unstable photon sphere)
\begin{align}
r_{PS}=\frac{M}{2} \left(3+\sqrt{9-8 \delta} \right). 
\end{align}
The expression for the photon shadow then has the form (\ref{eq:a9})
\begin{align}
R^2_{PS}=\frac{M^2}{8}\cdot\frac{\left(3+\sqrt{9 -8 \delta}\right)^4}{\left(3+\sqrt{9 -8 \delta} \right)-2 \delta}=27M^2\cdot\left(1-\frac{1 }{3}\cdot\delta-\frac{1}{27}\cdot\delta^2\right)+...
\end{align}
It is easy to check with any symbolic computation package that this result is the same as (19) in Ref. \cite{Vagnozzi:2022moj} and (21) in Ref. \cite{Vertogradov:2024dpa} for $M=1$.

\begin{figure}[tb]
\centering
\includegraphics[scale=0.7]{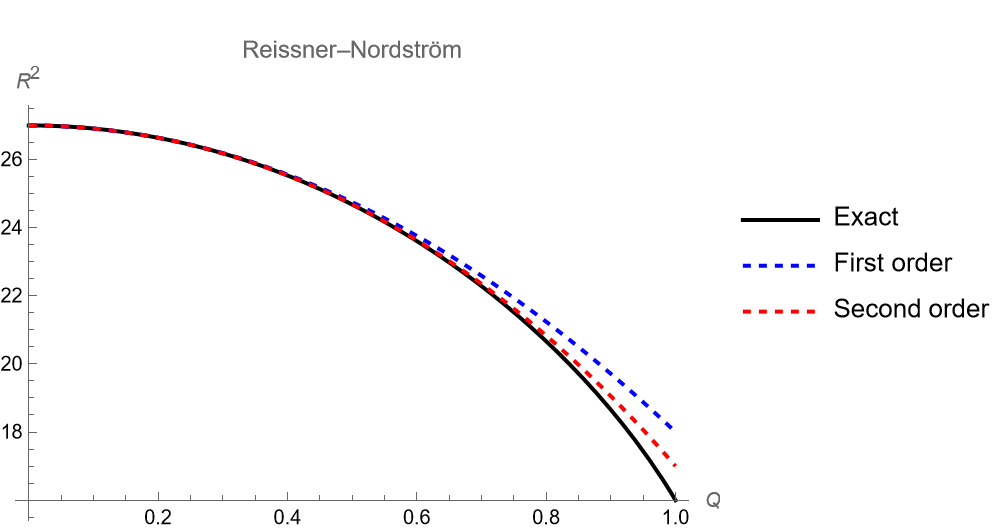} 
		\caption{Reissner-Nordström photon shadow. The dependence of the photon shadow square radius $R^2$ on charge $Q$ ($M=1$). The exact result, first-order approximation, second-order approximation are presented.}
		\label{fig:RN}
\end{figure}

\begin{figure}[tb]
\centering
\includegraphics[scale=0.7]{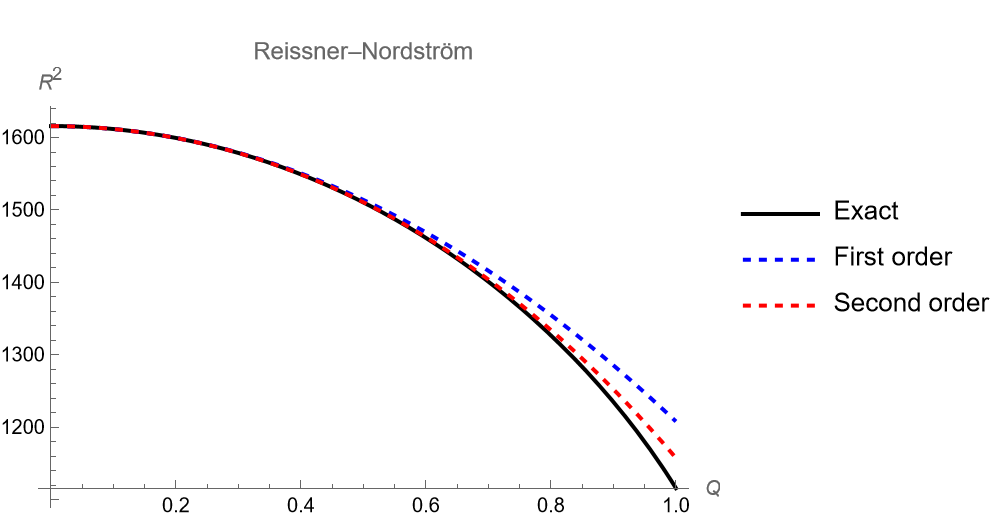} 
		\caption{Reissner-Nordström massive shadow $\epsilon=0.99$. The dependence of the massive shadow square radius $R^2$ on charge $Q$ for $\epsilon=0.99$ ($M=1$). The exact result, first-order approximation, second-order approximation (\ref{eq:delta_expansion}) are presented.}
		\label{fig:RN1}
\end{figure}

The general expression for a massive shadow can be obtained using the expansion (\ref{eq:b2}) and the first of the equations (\ref{eq:area_shadow}) and has the form
\begin{align} \label{eq:b7}
R^2=& 27M^2\cdot\Big[1+\frac{2}{3}\cdot \epsilon-\frac{1}{3}\cdot \delta+\frac{17}{27}\cdot\epsilon^2-\frac{5 }{27 }\cdot \delta \epsilon -\frac{1}{27 }\cdot \delta^2\Big]+...
\end{align}
Having in hand this series expansion of the general result, we can test our expressions (\ref{eq:a6}) or (\ref{eq:a7}). In our case
\begin{align}
\alpha_1(f)=f^{-2}, \quad \alpha_2(f)=0.
\end{align}
and from the general formulas (\ref{eq:b4}) and (\ref{eq:a6}) we get
\begin{align}
r=3 M +\frac{M}{3}\cdot \epsilon - \frac{2M}{3}\cdot \delta+...
\end{align}
and
\begin{align}
R^2= R^2_{Sch} \cdot\Big[1+\frac{2}{3}\cdot \epsilon-\frac{1}{3}\cdot\delta+\frac{17}{27}\cdot\epsilon^2-\frac{5 }{27 }\cdot \delta \epsilon -\frac{1}{27 }\cdot \delta^2\Big]+...
\end{align}
In this way we accurately reproduce the result (\ref{eq:b2}) and  (\ref{eq:b7}). We also provide a graphical illustration for photon shadow square radius as functions of $Q$ in Fig. \ref{fig:RN}. As can be seen, our approach gives a very good approximation even in the case of nearly extreme regime $Q=M$. This situation also holds for the massive case if we use the general formula  (\ref{eq:delta_expansion}) with $\epsilon=0.99$ as can be seen from the Fig. \ref{fig:RN1}.  

For illustrative purposes, we also apply the formalism (\ref{c3}) developed in the previous section to determine the charge $Q$. Applying Eq. (\ref{c3})  with $n=2$ we find
\begin{align}  \label{eq:c5}
Q^2=\frac{27(4-f)\chi-f^3}{9(4-f) \chi-f^{2}}\cdot M^2, \quad \chi=(R/R_{PS})^2,
\end{align}
for any $\epsilon$. 

Let's consider some numerical examples. Using the exact formula (\ref{eq:area_shadow}), we generate two sample measurements for $Q^2=0.01$ and $M=1$. First "experimental" example $\epsilon_{exp}=0.138611$, $\chi_{exp}=1.10646$. After substitution in (\ref{eq:c5}) we get
\begin{align}
Q^2_{exp}=0.010011.
\end{align}
Second example $\epsilon_{exp}=0.445219$,  $\chi_{exp}=1.51894$. After substitution in (\ref{eq:c5}) we get
\begin{align}
Q^2_{exp}=0.0100104.
\end{align}
These results agree with the true value with high accuracy. There is a small deviation due to the discarded higher orders of $Q^2$ which contribute to the exact formula (\ref{eq:area_shadow}) but it's even smaller than $Q^4$. 
In addition, the result is practically independent of the choice of $\epsilon$, which is the main criterion for the correct choice of the approximating model.     

\subsection{$r^{-n}$ metric}

The next example we want to consider reads as 
\begin{align} \label{eq:exa3}
\alpha=1-\frac{2M}{r} +\frac{M^n \delta}{r^n}, \quad \beta=r^2, \quad \delta=Q^n/M^n.
\end{align}
Like the Reissner-Nordström metric, it contains only one perturbing term of first order, but with a different power of $r$. Therefore, the defining equation for MPS in the general case have a degree higher than four and is not solvable in radicals. Therefore, we must use perturbation theory. Substituting (\ref{eq:exa3}) into the general formulas (\ref{eq:b4}) and (\ref{eq:a6}) we find 
\begin{align}
r=3 M +\frac{M}{3}\cdot \epsilon - M \cdot3^{1-n}\left(1+n/2\right)\cdot \delta+...
\end{align}
and
\begin{align}
R^2=& R^2_{Sch} \cdot\Big[1+\frac{2}{3}\cdot \epsilon-3^{1-n}\cdot \delta+\frac{17}{27}\cdot\epsilon^2+ 3^{-n-1} (n-7)\cdot \delta \epsilon\nonumber\\&-\frac{3^{1-2 n}}{4} \left(n^2+4 n-8\right)\cdot \delta^2\Big]+...
\end{align}
It is easy to check that for $n=2$ this result reduces to previous one. Interestingly, larger powers of $n$ lead to smaller perturbation $\propto 3^{-n}\delta$.  

Let us now apply the method (\ref{c3}) of determining the metric coefficients to this solution. This time we need to consider only the coefficient $c_i$ of the highest power $n$ to be non-zero. This leads to the 
\begin{align} \label{eq:c6}
Q^n=\frac{27(4-f)\chi-f^3}{3^{4-n}\left(4-f\right) \chi-f^{4-n}}\cdot M^n.
\end{align}
for any $\epsilon$. Consider a numerical example again. We choose $n=10$, $Q^{10}=0.01$ and $M=1$. First example $\epsilon_{exp}=0.256197$, $\chi_{exp}=1.22597$. After substitution in (\ref{eq:c6}) we get
\begin{align}
Q^{10}_{exp}=0.0100002.
\end{align}
Second example $\epsilon_{exp}=0.738405$,  $\chi_{exp}=2.76522$, we get
\begin{align}
Q^{10}_{exp}=0.0100001.
\end{align}
The error is smaller than in the case of Reissner-Nordström metric, which is due to the exponential nature of disturbances decrease noted earlier. 

Consider another toy example in this section:
\begin{align} \label{eq:exa4}
\alpha=1-\frac{2M}{r} +\delta \left(\frac{M^2}{r^2}+\frac{2M^3}{r^3}+\frac{3M^4}{r^4}\right), \quad \beta=r^2.
\end{align}
For this metric, we will try to reconstruct the coefficients $C_i$ from a set of massive shadows in accordance with the procedure (\ref{c3}). We choose $\delta=0.001$, $M=1$ and generate a set of shadow examples as before:
\begin{align} \label{eq:f5}
\epsilon^{exp}_j&=(0.138366, 0.256275, 0.357379),\\
\chi^{exp}_j&=(1.1062, 1.2261, 1.36214).
\end{align}
Then solving the general system (\ref{c3}) of linear inhomogeneous equations we find
\begin{align} \label{eq:f6}
C^{exp}_i=(0.00100089,0.00199377,0.00301871).
\end{align}
We see that for these coefficients the ratio of the original metric $1:2:3$ does indeed hold, although of course the error is greater than in the case of the Reissner-Nordström metric. For additional verification, we also consider another series of closer $\epsilon_j$
\begin{align} \label{eq:f3}
\epsilon^{exp}_j&=(0.0295166, 0.0580735, 0.0857096),\\
\chi^{exp}_j&=(1.02025, 1.04097, 1.0622).
\end{align}
We end up with
\begin{align} \label{eq:f4}
C^{exp}_i=(0.000980033,0.00205673,0.00295536).
\end{align}
Thus we can actually reconstruct the coefficients $C_i$ separately, which could not be done from the photon shadow in the first order of perturbation theory. Of course, from (\ref{eq:f3}) and (\ref{eq:f4}) we can see that the choice of closer values of the $\epsilon$ can lead to larger errors than for choice (\ref{eq:f5}) and (\ref{eq:f6}). We noted this fact earlier, in particular, for very close values $\epsilon_j$ we must use greater measurement precision or as in our case machine precision.

\subsection{Bardeen Black Hole}

\begin{figure}[tb]
\centering
\includegraphics[scale=0.7]{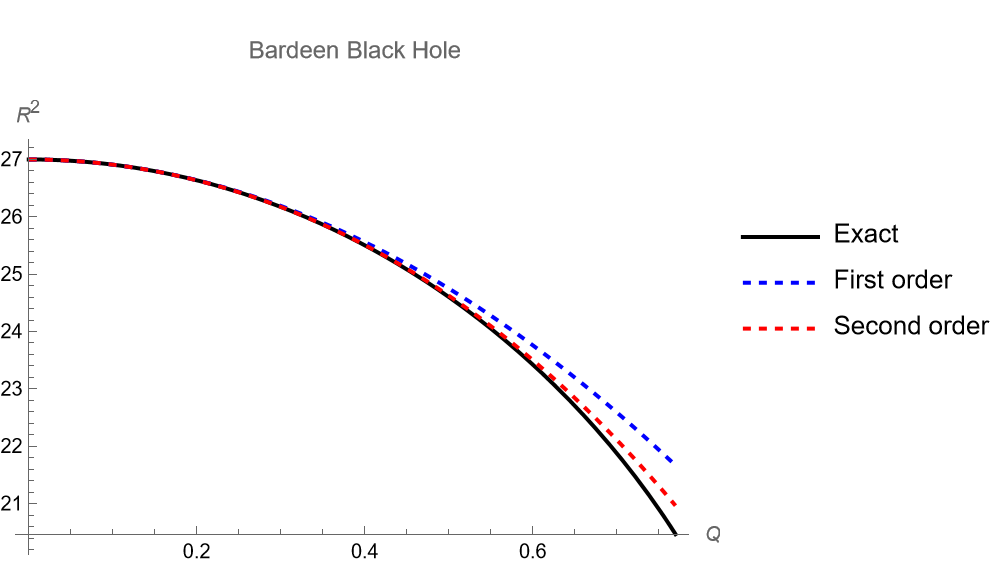} 
		\caption{Bardeen Black Hole photon shadow. The dependence of the photon shadow square radius $R^2$ on charge $Q\leq\sqrt{16/27}$ ($M=1$). The exact result, first-order approximation, second-order approximation are presented.}
		\label{fig:BBH}
\end{figure}

The metric functions for the Bardeen magnetically charged BH are given by \cite{Vagnozzi:2022moj}
\begin{align} \label{eq:exa5}
\alpha=1-\frac{2 M r^2}{\left(r^2+ M^2\delta \right)^{3/2}}, \quad \beta=r^2, \quad \delta=Q^2/M^2.
\end{align}
where $\delta$ characterizes a particular hair and satisfies the condition $\delta\leq 16/27$. Expanding (\ref{eq:exa5}) with respect to the parameter $\delta$, we find \cite{Vertogradov:2024dpa}
\begin{align} \label{eq:e1}
\alpha_1(f)=3 f^{-3}, \quad \alpha_2(f)=-\frac{15}{4}\cdot f^{-5}.
\end{align}
Substituting (\ref{eq:e1}) into the general formulas (\ref{eq:b4}) and (\ref{eq:a6}) we obtain for MPS radius
\begin{align}
r=3 M +\frac{M}{3}\cdot \epsilon - \frac{5M}{6}\cdot \delta+...
\end{align}
and for massive shadow radius
\begin{align}
R^2= R^2_{Sch} \cdot\Big[1+\frac{2}{3}\cdot \epsilon-\frac{1}{3}\cdot\delta+\frac{17}{27}\cdot\epsilon^2-\frac{4 }{27 }\cdot \delta \epsilon -\frac{2}{27 }\cdot \delta^2\Big]+...
\end{align}
Thus, in the first order the shadow behaves  the same as in the Reissner-Nordström metric, and the deviation can only be detected in the second order, which allows us to classify these solutions as simulating. We present a comparison of the exact and perturbative results in Fig. \ref{fig:BBH}. As in the case of the Reissner-Nordström metric, the approximation works well almost to the extreme regime $Q^2=16M^2/27$. 

\subsection{Magnetically charged Einstein-Euler-Heisenberg (EEH) Black Hole} 

Magnetically charged BHs arising from Einstein-Bronnikov NLED reads as \cite{Vagnozzi:2022moj} 
\begin{align}
\alpha=1-\frac{2M}{r}+\frac{M^2\delta}{r^2}-\frac{2\mu/M^2}{5}\cdot\frac{M^6\delta^2}{r^6}, \quad \beta=r^2, \quad \delta=Q^2/M^2.
\end{align}
where $Q$ is the BH magnetic charge and characterizes a specific hair, and $\mu$ is the NLED coupling. In our notation we find
\begin{align} \label{eq:exak1}
\alpha_1(f)=f^{-2}, \quad \alpha_2(f)=-\frac{2}{5}\cdot\mu/M^2\cdot f^{-6}.
\end{align}
Substituting (\ref{eq:exak1}) into the general formulas (\ref{eq:b4}) and (\ref{eq:a6}) we find
\begin{align}
r=3 M +\frac{M}{3}\cdot \epsilon - \frac{2M}{3}\cdot \delta+...
\end{align}
and
\begin{align}
R^2= R^2_{Sch} \cdot\Big[1+\frac{2}{3}\cdot \epsilon-\frac{1}{3}\cdot\delta+\frac{17}{27}\cdot\epsilon^2-\frac{5}{27 }\cdot \delta \epsilon -\frac{1}{27 }\cdot\left(1-\frac{2}{45}\cdot\mu/M^2\right)\cdot \delta^2\Big]+...
\end{align}
As in the previous case, we find that in the first order the shadow behaves absolutely identically to the Reissner-Nordström metric and only in the second order we can observe a slight deviation. Note also the presence of a critical value $\mu_{c}=45M^2/2$ which separates two alternative second order shadow behaviors. However, such a value is not achieved because  $\mu/M^2=0.3$ is approximately the largest allowed coupling before the
perturbative approach of the theory around the Maxwell
Lagrangian ceases to be meaningful \cite{Vagnozzi:2022moj}. 

\subsection{Fisher-Janis-Newman-Winicour metric}

\begin{figure}[tb]
\centering
\includegraphics[scale=0.7]{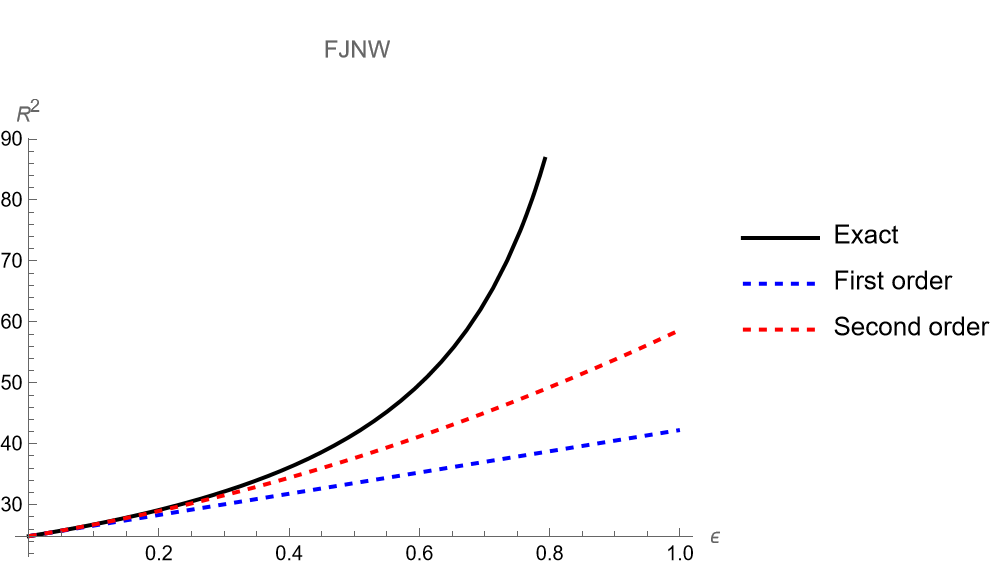} 
		\caption{FJNW massive shadow. The dependence of the massive shadow square radius $R^2$ on $\epsilon$ with $\sigma= 3/4$ ($M=1$). The exact result and the first and the  second-order approximations are presented.}
		\label{fig:FJNW}
\end{figure}

The metric functions for FJNW solution is given by \cite{Fisher:1948yn,Janis:1968zz,Abdolrahimi:2009dc} 
\begin{align}
\alpha=\left(1-\frac{2 M}{r \sigma }\right)^{\sigma}, \quad\beta=r^2 \left(1-\frac{2 M}{r \sigma }\right)^{1-\sigma}.
\end{align}
For this metric we can put $\delta=1-\sigma$, however in this notation a different coordinate $r$ is used, since $\beta\neq r^2$. So, we either have to use more general formulas than in the previous examples or apply the formula (\ref{eq:epsilon_expansion}) since the photon sphere is easily defined. Indeed photon sphere reads as $r_{PS}=\frac{2 \sigma +1 }{\sigma }\cdot M$. Then massive particles surface expansion (\ref{eq:epsilon_expansion_r}) reads as
     \begin{align}
    r = \frac{2 \sigma +1 }{\sigma }\cdot M + \left(\frac{2 \sigma -1}{2 \sigma +1}\right)^{\sigma}\cdot  M \cdot \epsilon +O(\epsilon).
    \end{align}
    Photon shadow radius (\ref{eq:a9}) reads as
    \begin{align}
    R^2_{PS}=\frac{(2 \sigma-1)^2 }{\sigma^2}\cdot\left(\frac{2 \sigma+1}{2 \sigma-1}\right)^{2 \sigma+1}\cdot M^2.
    \end{align}
    And finally, the massive shadow expansion over the photon sphere (\ref{eq:epsilon_expansion}) is
    \begin{align}
    R^2=R^2_{PS}\cdot \left[1+\left\{1-\left(\frac{2 \sigma-1}{2 \sigma+1}\right)^{\sigma}\right\}\cdot(\epsilon+ \epsilon^2)-\frac{\sigma^2 }{ (2\sigma+1)^2}\cdot\left(\frac{2 \sigma-1}{2 \sigma+1}\right)^{2 \sigma-1}\cdot \epsilon^2\right] +O(\epsilon^2).  
\end{align}
We illustrate this approximation in Fig. \ref{fig:FJNW}.

For uniformity of results, let's also expand this result in terms of $\delta=1-\sigma$. We get 
\begin{align}
r=3 M+\frac{M}{3}\cdot\epsilon+M\cdot\delta+ ...
\end{align}
and
\begin{align}
R^2=& 27M^2\cdot\Big[1+\frac{2}{3}\cdot \epsilon-2(\ln3-1)\cdot \delta+\frac{17}{27}\cdot\epsilon^2-\left(\frac{5 \ln 3}{3}-\frac{16}{9}\right)\cdot \delta \epsilon \\&-\left(4 \ln 3 -2 \ln^23-\frac{5}{3}\right)\cdot \delta^2\Big]+...
\end{align}
Note that the shadow, as in previous cases, shrinks with increasing $\delta$, however, the radius of the MPS is growing.

\subsection{Results table}

We collect all the results in a unified form 
\begin{align}
r=3 M+\frac{M}{3}\cdot\epsilon-M \cdot A\cdot\delta+O(\epsilon,\delta),
\end{align}
and
\begin{align} 
R^2=& 27 M^2 \cdot\Big[1+\frac{2}{3}\cdot \epsilon-B\cdot\delta+\frac{17}{27}\cdot\epsilon^2-C\cdot\epsilon \delta-D\cdot \delta^2\Big]+O(\epsilon^2,\delta^2,\epsilon \delta),
\end{align}
where $A$, $B$, $C$, $D$ are presented in the table \ref{tb1}. The table also provides the following additional examples ($\beta=r^2$):

\begin{itemize}
    \item Hayward regular BH \cite{Hayward:2005gi}: 
    \begin{align} 
    \alpha=1-\frac{2 M r^2}{2 \delta  M^3+r^3}=1-\frac{2 M}{r}+\frac{4 \delta  M^4}{r^4}-\frac{8 \delta ^2 M^7}{r^7}+...
    \end{align}
    \item  Ghosh-Kumar BH \cite{Ghosh:2021clx}:
    \begin{align} 
    \alpha=1-\frac{2 M}{\sqrt{\delta  M^2+r^2}}=1-\frac{2 M}{r}+\frac{\delta  M^3}{r^3}-\frac{3 \delta ^2 M^5}{4 r^5}+...
    \end{align}
     \item Ghosh-Culetu-Simpson-Visser (GCSV) BH \cite{Ghosh:2014pba}
     \begin{align} 
    \alpha=1-\frac{2 M \exp \left(-\frac{1}{2}\frac{\delta  M}{r}\right)}{r}=1-\frac{2 M}{r}+\frac{\delta  M^2}{r^2}-\frac{\delta ^2 M^3}{4 r^3}+...
    \end{align}
    \item Kazakov-Solodukhin regular BH \cite{Kazakov:1993ha}:
     \begin{align} 
    \alpha=\frac{\sqrt{r^2-\delta  M^2}}{r}-\frac{2 M}{r}=1-\frac{2 M}{r}-\frac{\delta  M^2}{2 r^2}-\frac{\delta ^2 M^4}{8 r^4}+...
     \end{align}
\end{itemize}

Note that Reissner-Nordström, Bardeen, GCSV and EEH are not distinguishable by the shadow behavior in the first order of perturbation theory (column $B$) while all the others have unique behavior already. In the second order, all solutions are distinguishable. Therefore, our choice of expansion including the second order appears to be the most motivated.

\begin{center}
\begin{longtable}{ |c||c||c|c|c| }
\caption{Perturbation coefficients. \label{tb1}}\\
 \hline
 Metric & $A$ & $B$ & $C$ & $D$ \\ 
 \hline\hline
 Schwarzschild & $0$ & $0$& $0$& $0$\\ 
 \hline
 Reissner-Nordström & $2/3$ & $1/3$& $5/27$& $1/27$\\ 
 \hline
 Bardeen & $5/6$ & $1/3$& $4/27$& $2/27$\\ 
 \hline
  Hayward & $4/9$ & $4/27$& $4/81$& $8/243$\\ 
  \hline
  Ghosh-Kumar & $5/18$ & $1/9$& $4/81$& $1/243$\\ 
  \hline
GCSV & $2/3$ & $1/3$& $5/27$& $1/108$\\ 
  \hline
  Kazakov-Solodukhin & $-1/3$ & $-1/6$& $-5/54$& $1/216$\\ 
  \hline
EEH  & $2/3$ & $1/3$& $5/27$& $1/27\cdot\left(1-2/45\cdot\mu/M^2\right)$\\ 
  \hline
  FJNW  & $-1$ & $2\ln3-2$ & $5/3 \ln 3-16/9$ & $4 \ln 3 -2 \ln^23-5/3$\\ 
   \hline
    $r^{-n}$ metric & $3^{1-n}\left(1+n/2\right)$ & $3^{1-n}$ & $3^{-n-1} (7-n)$ & $3^{1-2 n} \cdot2^{-2}\left(n^2+4 n-8\right)$\\ 
 \hline
\end{longtable}
\end{center}

\section{Conclusion} \label{sec:conclusion}
We constructed two-parameter expansions to second order of the massive particle spheres and shadow boundaries seen by a distant observer in terms of the mass-energy ratio and the spacetime deformation parameters around the standard picture corresponding to a Schwarzschild black hole.
This may help detect deviations of the shadow from the Schwarzschild pattern due to additional parameters of the black hole, such as charge or some other structural parameter of the solution of the Einstein or modified gravity equations. More precisely, we developed and tested a simple method for determining the metric parameters from a set of massive shadows using a system of linear inhomogeneous equations. It is noteworthy that this information cannot be obtained only from the photon shadow. This approach does not require information such as the distance to the black hole or its mass, but only potentially observable images on Earth. Although experimental observation of such shadows is difficult to implement, this formalism can be easily generalized to the case of photons in plasma \cite{Perlick:2015vta,Perlick:2023znh,Kobialko:2023qzo}, which clearly brings the problem back into the direct experimental context. We also examined a number of gravitational models and found that Reissner-Nordström, Bardeen, GCSV and EEH black holes are not distinguishable by their shadows  in the first order of perturbation theory while in the second order all models are distinguishable. These results are directly applicable to photon shadows also. This can be helpful in reading off extra information from experimentally observed images of black holes and other compact objects.

\begin{acknowledgments}
The authors thanks Igor Bogush for useful suggestions
and discussions. This work was supported by Russian Science Foundation under Contract No. 23-22-00424.
\end{acknowledgments}

\bibliography{main}

\end{document}